\documentclass{article}
\usepackage{graphicx,graphics}

\newcommand{\beq}{\begin{equation}}
\newcommand{\eeq}{\end{equation}}
\newcommand{\bea}{\begin{eqnarray}}
\newcommand{\eea}{\end{eqnarray}}

\begin{document}
\centerline{\bf \large The relative significance of the $h$-index}
\medskip
\medskip
%\titlerunning{to be set}

\centerline{\bf \large H.C.\ Spruit}
%\authorrunning{H.C.\ Spruit}

%\offprints{\\ H.\ Spruit, \email{henk@mpa-garching.mpg.de}}

\medskip
\centerline{\large  Max-Planck-Institut f\"{u}r Astrophysik,}\vskip 0.2\baselineskip
\centerline{\large   Karl-Schwarzschild-Str.\ 1,}\vskip 0.2\baselineskip
\centerline{\large    85748 Garching, Germany}
\medskip
%\date{\today}
\abstract{\small Use of the Hirsch-index ($h$) as measure of an author's visibility in the scientific literature has become popular as an alternative to a gross measure like total citations (c). I show that, at least in astrophysics, $h$ correlates tightly with overall citations. The mean relation is $h=0.5(\sqrt c+1)$. Outliers are few and not too far from the mean, especially if  `normalized' ADS citations are used for $c$ and $h$. Whatever the theoretical reasoning behind it, the Hirsch index in practice does not appear to measure something significantly new. 

%\keywords{citations -- $h$-index}\ehm
}

%\maketitle

\vskip 1\baselineskip
{\bf  Citation metrics}
\vskip 0.5\baselineskip

The Hirsch index, a quantitative measure of  an authors' status in the scientific literature, is defined as the paper number $n$ in his list of publications (ranked by citations, highest first) where $n$ equals the number of citations. (In NASA's ADS, click on `sort by normalized citations' and scroll the results down to the last paper of which the rank is less or equal to the number of citations listed for it). It appears to have become fashionable as a more meaningful metric than, or as an alternative to, the simple metric of raw citation rates. It has been implemented also for quantifying the scientific impact of observing facilities (Grothkopf et al. 2007). 

The invention of new figures of merit using citation rates implies an attempt to find optimal metrics for an intuitive notion of quality (`A better than B'). Reflection on the way in which these statistics are handled in practice, however, shows that focus on the accuracy or biases of different metrics in abstracto is somewhat misguided. The uncertainty in applying such metrics is related more to the purpose they are to be used for. Unreliable work, for example, can have a high citation rate but this will be discounted accordingly if reliability is more important, depending on the purpose of an evaluation. High visibility in the popular media, on the other hand, can trump reliability, teaching skills can trump citation rates, etc. The substitution of  gross citations  by a new index like $h$ as an abstract measure of `quality' or impact tends to obscure this more basic issue. 

For the average astrophysical career, $h$ scales roughly linearly with the author's career age (time since PhD). Like total citations, it can therefore not be used for comparison between authors unless compensation is made for differences in career age. The total number of all citations in the reference lists of the papers published in the main journals has increased at a fairly constant rate of 5\% per year since the 1960's. The starting date of one's career thus also makes a significant difference: younger scientists should be expected to collect citations faster than their older colleagues did at the same stage in their career. [Similarly: the yearly increase in your citation rate has to be discounted somewhat due to this increase, if citations are to be interpreted as your relative impact in the community]. The use of citation numbers, even if  understood as measuring a more neutral descriptive like visibility rather than quality, is therefore not meaningful unless such factors are taken into account.

\medskip
\begin{figure}[ht]
\includegraphics[width=1.0\hsize]{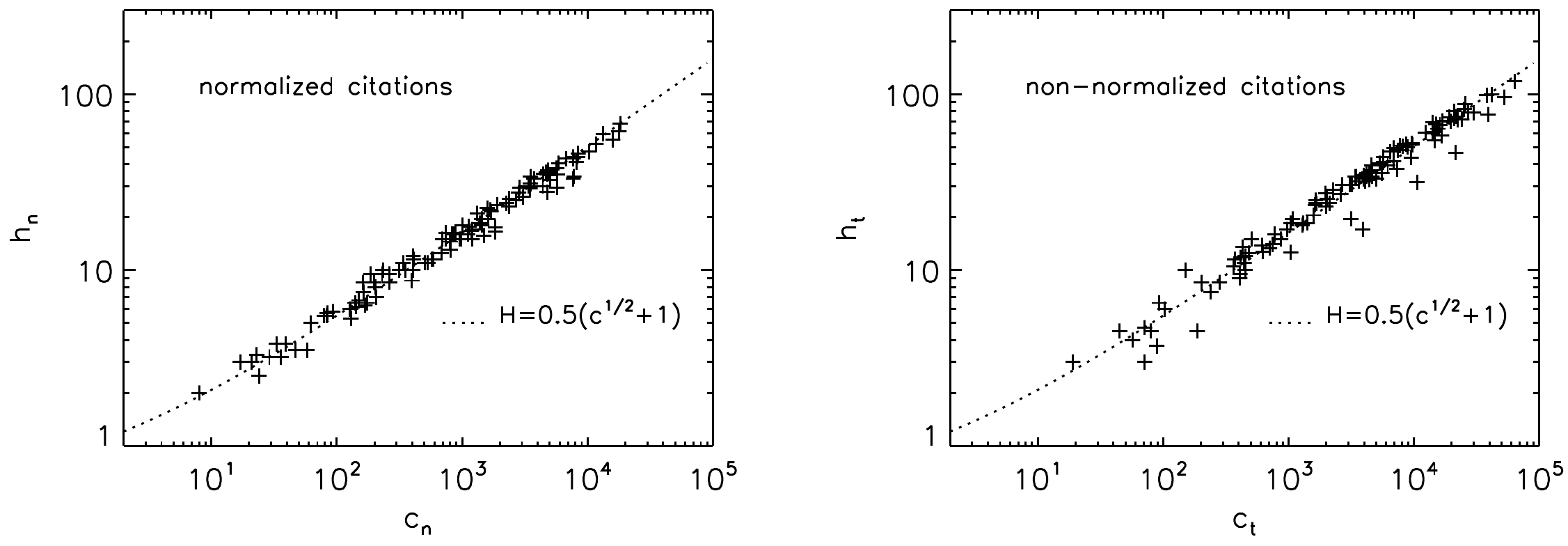}
\caption{Correlation between Hirsch-index ($h$) and citations ($c$) using ADS normalized citations (left) and ADS non-normalized citations (right).}
\end{figure}

{\bf $h$-index vs. total citations}
\vskip 0.5\baselineskip
\noindent
Figure 1 shows $h$-indices for a sample of astrophysicists, using citation numbers from ADS\footnote{These are more reliable than numbers from commercial products since they take into account more information specific to the field of astrophysics.}. Two different versions are shown, the $h$-index $h_{\rm t}$ using total citations $c_{\rm t}$, and the index $h_{\rm n}$ based on ADS `normalized' citations $c_{\rm n}$. The latter weighs the citations by the number of co-authors on the paper cited. This is appropriate if the thing  to be measured is not the impact of a paper but of an individual author, and makes it possible to compare authors working in collaborations of different size. The sample (113 points) has been compiled through a non-random selection process. Most points represent names that occurred to me while scanning the daily arXiv astro-ph abstracts. On average, these score higher than the mean. To get better statistics at the low end, I added a number of authors suspected to score lower on the basis of personal acquaintance, plus a number of junior authors of papers with better known senior coauthors. The distribution of points along the curve is thus not representative. A good fit to the data is 
\begin{equation}h=0.5(\sqrt c+1).\end{equation} (The 1 in the bracket has been added so the curve also fits the hypothetical author with just 1 citation). The square-root dependence was noted by Hirsch (2005), the equivalence of the $h$-index with total citations was emphasized already  by Nielsen (2008), both for samples of physicists (see also Petersen et al. 2011). The scatter around this curve is 11\% rms for normalized citations and 18\% for non-normalized citations. Scatter at this level is likely to be well within the uncertainties associated with any relevant evaluation process. The increased scatter when using non-normalized citations is due to authors publishing in large collaborations (points to the right of the bulk).  

Debates about the virtues of the $h$-index\footnote{The main reason for the popularity of the $h$-index is probably the organization of the ISI citation index that is used for most countings. Adding up the total citations  (via the ISI web interface WOS) is laborious  (ISI can provide such numbers at an additional charge).  An $h$-index is found much more easily.} have centered on theoretical arguments about possible biases intrinsic to different metrics (e.g.\ Hirsch 2005, Redner 2010, Waltman \& van Eck  2011, Waaijers 2011). In practice, the Hirsch index does not appear to measure anything significantly different from overall citations, however, at least not for the present sample of astrophysicists (nor for physics in general, Nielsen 2008). 

\vskip 1\baselineskip
{\bf References}
\vskip 0.5\baselineskip
\noindent
Grothkopf, U., Melo, C., Erdmann, C., Kaufer, A., \& Leibundgut, B.\ \\ \indent 2007, The Messenger, 128, 62 (2007Msngr.128...62G)\\
Hirsch J. E. 2005, Proc.\ Nat.\ Acad.\ Sci.\ 102, 16569 (arXiv:physics/0508025) \\
Nielsen, M., 2008, \\ \indent http://michaelnielsen.org/blog/why-the-h-index-is-virtually-no-use/\\
Petersen, A.M., Stanley, H.E., \& Succi, S., 2011, Scientific Reports, 1, \\ \indent a181. ~doi:10.1038/srep00181\\
Redner, S., 2010, J. Stat. Mech. L03005 (arXiv:1002.0878v1\\ \indent [physics.data-an])\\
Waaijers, L. ,2011, arXiv:1109.5520v1 [physics.soc-ph]\\
Waltman, L., van Eck, N.J., 2011, arXiv:1108.3901v1 [cs.DL]\\

\end{document}